\documentclass[aps,pre,reprint,superscriptaddress]{revtex4-1} 
\usepackage{amsmath,amssymb,graphicx,bm,natbib} 
\pdfoutput=1
\DeclareGraphicsExtensions{.pdf} 

\begin{document} 

\title{Spatially embedded growing small-world networks} 
\author{Ari Zitin} 
\affiliation{Institute for Research in Electronics and Applied Physics, 
             University of Maryland, College Park, Maryland 20742, USA} 
\author{Alex Gorowara} 
\affiliation{Institute for Research in Electronics and Applied Physics, 
             University of Maryland, College Park, Maryland 20742, USA} 
\author{Shane Squires} 
\affiliation{Institute for Research in Electronics and Applied Physics, 
             University of Maryland, College Park, Maryland 20742, USA} 
\affiliation{Department of Physics, University of Maryland, College Park, 
             Maryland 20742, USA} 
\author{Mark Herrera} 
\affiliation{Institute for Research in Electronics and Applied Physics, 
             University of Maryland, College Park, Maryland 20742, USA} 
\affiliation{Department of Physics, University of Maryland, College Park, 
             Maryland 20742, USA} 
\affiliation{Heron Systems Inc., 2121 Eisenhower Rd., Alexandria, VA 22314, USA} 
\author{Thomas M.\ Antonsen} 
\affiliation{Institute for Research in Electronics and Applied Physics, 
             University of Maryland, College Park, Maryland 20742, USA} 
\affiliation{Department of Physics, University of Maryland, College Park, 
             Maryland 20742, USA} 
\affiliation{Department of Electrical and Computer Engineering, University 
             of Maryland, College Park, Maryland 20742, USA} 
\author{Michelle Girvan} 
\affiliation{Institute for Research in Electronics and Applied Physics, 
             University of Maryland, College Park, Maryland 20742, USA} 
\affiliation{Department of Physics, University of Maryland, College Park, 
             Maryland 20742, USA} 
\affiliation{Institute for Physical Science and Technology, University of 
             Maryland, College Park, Maryland 20742, USA} 
\author{Edward Ott} 
\affiliation{Institute for Research in Electronics and Applied Physics, 
             University of Maryland, College Park, Maryland 20742, USA} 
\affiliation{Department of Physics, University of Maryland, College Park, 
             Maryland 20742, USA} 
\affiliation{Department of Electrical and Computer Engineering, University 
             of Maryland, College Park, Maryland 20742, USA} 

\date{\today}

\begin{abstract} 
Networks in nature are often formed within a spatial domain in a 
dynamical manner, gaining links and nodes as they develop over time.  
We propose a class of spatially-based growing network models and 
investigate the relationship between the resulting statistical network 
properties and the dimension and topology of the space in which the 
networks are embedded.  In particular, we consider models in which 
nodes are placed one by one in random locations in space, with each 
such placement followed by configuration relaxation toward uniform 
node density, and connection of the new node with spatially nearby 
nodes.  We find that such growth processes naturally result in 
networks with small-world features, including a short characteristic 
path length and nonzero clustering.  These properties do not appear 
to depend strongly on the topology of the embedding space, but 
do depend strongly on its dimension; higher-dimensional spaces 
result in shorter path lengths but less clustering.  
\end{abstract} 


\maketitle

\section{Introduction} 

One fascinating property of many real-world networks is that they 
are often ``small worlds'' in the sense that they have both {\em 
short average path length} and {\em high clustering} \cite{milgram,%
wsnat,newmanreview}.  The shortest path length between two nodes is 
the smallest number of links in the path connecting that pair of nodes, 
and the average path length is the average of this value over all node 
pairs in the network.  It is regarded as short if it grows very slowly 
with network size.  To quantify the clustering of an undirected network, 
we use the clustering coefficient, which is defined as three times the 
number of triangles in the network divided by the number of link pairs 
that share a common node \cite{newmanrandom}.  In networks with high 
clustering, if two nodes are both neighbors of a third node, they are 
also likely to be connected to one another.  A variety of real-world 
networks, from social networks to neuronal networks, exhibit the small-%
world property, and this has fundamental consequences for dynamical 
processes such as spread of information or disease \cite{wsnat}.  

Networks with spatial constraints typically have geographically short-%
range edges, and it is thus relevant that both the original Watts-Strogatz 
small-world model \cite{wsnat} and many real networks with the small-%
world property are embedded in physical space.  For example, the Internet, 
a network of routers connected via cables, is essentially embedded on 
the two-dimensional surface of the Earth and tends to have mostly local 
links, presumably due to the cost of wiring \cite{internetgeography}.  
This has led many researchers to consider network models with spatial 
embedding \cite{ozik2004,przuljgeo,hermannspace,bullockspatial,guan1D,%
zhang2006,zhang2007,neuronembedding,barthelemy,vazquez2002}.  Work on 
this topic has revealed that small-world properties are found in a 
variety of spatially embedded networks, including networks of neurons, 
power grids, and social interactions \cite{wsnat,swbrain}.  

Two other key aspects of many real-world networks are that they 
grow with time (new nodes are added), and that nodes may move in 
space.  For example, new people may join social networks with time, 
and friendships typically form between people who live near one 
another, but people may also move to new locations.  Although some 
studies have considered dynamically growing networks, they frequently 
assume that nodes remain fixed in their initial positions 
\cite{bullockspatial,guan1D,przuljgeo} or consider growing networks 
which are not embedded in space \cite{bornholdt02,reallyrandom}.  

In Ref.~\cite{ozik2004}, Ozik et al.\ considered a model which 
incorporates both a growing number of nodes and node movement.  
In this model, nodes are placed randomly on the circumference 
of a circle, but undergo small displacements to maintain a 
constant density over time.  Each node initially forms links 
only to its geographic neighbors, but, due to growth, these 
links can subsequently be stretched in length, becoming long-%
range.  Due to the emergence of these long-range links, this 
model also generates networks with the small-world property, 
but in this case it is a consequence of the growth process, 
rather than the spontaneous formation of long-range edges.  
However, since the physical properties of typical spatial 
systems typically depend on the dimension of the embedding 
space, the main limitation of Ref.~\cite{ozik2004} is that 
only a one-dimensional space (the circle) is treated.  Thus, 
in this paper, we generalize the model of \cite{ozik2004} by 
introducing and analyzing a class of growing undirected network 
models that have spatially constrained nodes able to move about 
in an embedding space of arbitrary dimension.  (We note that, 
for real applications, dimensions two and three are commonly 
most relevant.)

\section{The Circle Network Model} 

In Ref.~\cite{ozik2004}, the authors presented a model, henceforth 
referred to as the Circle Network Model, which considers an undirected 
network which initially has $m+1$ uniformly separated, all-to-all 
connected nodes on the circumference of a circle.  At each discrete 
growth step the network is grown according to the following rules: 
\begin{enumerate} 
\item A new node is placed at a randomly selected point on the 
      circumference of the circle.  
\item The new node is linked to its $m$ nearest neighbors ($m$ is 
      even in Ref.\ \cite{ozik2004}).  
\item Preserving node positional ordering, the nodes are repositioned 
      to make the nearest-neighbor distances uniform.  
\item Steps (1-3) are repeated until the network has $N$ nodes.  
\end{enumerate} 
It has been shown that this growth model leads to a small-world 
network with an exponentially decaying degree distribution \cite{%
ozik2004}.  The original goal of the circle network model was to 
explore the effect that local geographic attachment has on the 
growth of networks and was partially motivated by the growth of 
biological (e.g., neuronal) networks as an organism develops from 
an embryo.  In the present paper, we extend this analysis by 
considering networks growing by geographic attachment preference 
in more general spaces.  

We define a network growth procedure to yield the small-world property 
if, as $N \to \infty$, (i) the average degree $\langle k \rangle$ 
of a node approaches a finite value; (ii) the characteristic graph 
path length $\ell$, the average value of the smallest number of links 
in a path joining a pair of randomly chosen nodes, does not grow 
with $N$ faster than $\log N$, as in an Erd\H{o}s-R\'enyi random 
network \cite{randomgraph,newmanrandom}; and (iii) the clustering 
coefficient $C$, the fraction of connected network triples which 
are also triangles, approaches a nonzero constant with increasing 
$N$ \footnote{In \cite{ozik2004} an alternate definition of the 
clustering coefficient was used.  Specifically, the local clustering 
$C_i$ of each node $i$ is $C_i = q_i/[\frac{1}{2} k_i (k_i-1)]$, 
where $q_i$ is the number of links between the $k_i$ neighbors 
of node $i$, and the global network clustering is the average of 
$C_i$ over $i$.}.  The circle network model exhibits all three 
properties.  

\textit{Degree distribution:} The degree distribution $H(k)$ is the 
probability that a randomly selected node has $k$ network connections.  
For large $N$, the degree distribution of the circle network model 
approaches 
\begin{equation}\label{eq:degdist} 
H(k) = \frac{1}{m+1}\left(\frac{m}{m+1}\right)^{k-m} 
\end{equation} 
for $k \geq m$, and $H(k) = 0$ for $k < m$ \cite{ozik2004}.  Since 
the number of new links added each time a new node is added is $m$, 
Eq.~\eqref{eq:degdist} yields the result that the average node degree 
$\langle k \rangle$ is $2m$, satisfying the criterion (i) for the 
small-world property.  

\textit{Characteristic path length:} In the circle network model, 
simulation results show that $\ell \sim \log N$, satisfying criterion 
(ii).  This may be explained intuitively by noting that as new nodes 
are added, they push apart the older connected nodes, lengthening the 
spatial distance traversed by older edges.  These older nodes can 
then have geographically long links, thus dramatically decreasing 
the shortest graph path length between any given pair of nodes.  

\textit{Clustering coefficient:} For the circle network model, it was 
shown that the clustering coefficient approaches a constant, positive, 
$m$-dependent value as $N \to \infty$, satisfying criterion (iii).

\section{Generalizing the Circle Network Model}

Like the Watts-Strogatz model, the circle network model may be described 
as a one-dimensional ring model in which connections are initially formed 
with $m$ nearest neighbors.  However, in the circle network model, long-%
distance edges do not form spontaneously, but are a natural result of the 
dynamics of network growth.  Moreover, the circle network model naturally 
raises the question of whether networks grown in higher-dimensional spaces 
exhibit similar properties.  A primary goal of this article is to address 
this question.  

In what follows, we introduce two models that generalize the model of 
Ref.\ \cite{ozik2004} to higher dimensionality (Secs.\ \ref{sec:spheremodel} 
and \ref{sec:ppmodel}), and then present our results from analysis of 
these models (Secs.\ \ref{sec:degdist} and \ref{sec:pathclust}).  Our 
main results are as follows.
\begin{enumerate}
\item[($i$)] The coupling of network growth with local geographical 
attachment leads to small-world networks independent of the dimension 
of the underlying space.
\item[($ii$)] The nodal degree distribution (Fig.\ \ref{fig:degdist}) 
and age-degree relationship (Fig.\ \ref{fig:degage}) are independent 
of dimension as in ($i$).
\item[($iii$)] The path length $\ell$ scales as $\log N$ with a 
coefficient that decreases with dimension (Fig.\ \ref{fig:pathlength}) 
for fixed average degree.
\item[($iv$)] The clustering coefficient $C$ approaches a finite 
asymptotic value with increasing $N$ (Fig.\ \ref{fig:clustN}) and 
this asymptotic value decreases with increasing dimension $d$ of the 
embedding space.
\item[($v$)] All of our results above appear to be independent of 
the global topology of the embedding space.
\end{enumerate}

\section{The Sphere Network Model\label{sec:spheremodel}} 

One natural generalization of embedding nodes on the one-dimensional 
circumference of a circle is to embed them on the two-dimensional 
surface of a sphere, or more generally on the $d$-dimensional surface 
of a hypersphere.  The case $d=1$ corresponds to the circle network 
model.  However, although it is trivial to arrange $N$ points along 
the circumference of a circle with uniform spacing, the analogous 
procedure is less well-defined on higher dimensional surfaces.  One 
way to generalize the arrangement procedure is to consider nodes to 
act like point charges and to move them to a minimum electrostatic 
energy equilibrium configuration.  The problem of finding the equilibrium 
configuration of point charges on the surface of a sphere dates back 
to 1904 when J.~J.~Thomson introduced his model of the atom, and the 
problem of obtaining such an equilibrium is sometimes referred to as 
the ``Thomson problem'' \cite{thomson1904}.  A related ``generalized 
Thomson problem'' assumes that the force between ``charges'' is 
proportional to $r^{-\alpha}$, where $r$ is the distance between 
charges, with $\alpha$ not necessarily equal to the Coulomb value, 
$\alpha = d$ \cite{nelson}.  For reasons discussed in Sec.\ 
\ref{sec:ppmodel}, we use the value $\alpha = d-1$ in simulations.  

Using the generalized Thomson problem as a guide, we develop a 
generalization of the circle network model, which we call the Sphere 
Network Model, as follows.  We model the nodes as point charges 
confined to a unit spherical surface of dimension $d$.  We successively 
add a new node onto the surface at random with uniform probability 
density per unit area and then add links to connect it to its $m$ 
nearest neighbors, where distance is defined as the shortest great 
circle path along the surface of the sphere between two nodes.  Next 
we relax the node positions to minimize the potential energy of the 
configuration using a gradient descent procedure, 
\begin{align} 
\label{eq:spheregrad} 
\frac{\text{d}\bm{x}_i}{\text{d}t} &= P[\bm{F}_i], \\ 
\label{eq:force} 
\bm{F}_i &= \sum_{j \neq i} \frac{\phantom{|}\bm{x}_i - %
            \bm{x}_j\phantom{|^d}}{|\bm{x}_i - %
            \bm{x}_j|^d}, 
\end{align} 
where $\bm{x}_i$ is the $(d+1)$ dimensional position vector of node 
$i$, $|\bm{x}_i|=1$ for all $i$, and $P[\cdot]$ denotes projection 
onto the $d$-dimensional surface of the sphere.  We note that, as 
new nodes are added, this procedure tends to yield a local energy 
minimum, as opposed to the global minimum (for some applications, 
such as modeling biological network growth, the identification of 
local rather global minima might be viewed as more appropriate.)  
Note that, for large $N$, the repulsive interaction ensures that 
the points are distributed approximately uniformly on the surface 
of the sphere.

\section{The Plum Pudding Network Model\label{sec:ppmodel}} 

The sphere network model described above has the topological feature 
that the geographical embedding region does not have any boundary, 
which allows us to find a nearly-uniform distribution of nodes by 
imagining them to be identical charges with repulsive interactions.  
We have also tested another model with a different topology having 
a boundary and using a different mechanism to encourage uniform 
distribution of nodes.  We call this second model the Plum Pudding 
Network Model after Thomson's famous model of the atom \cite{thomson1904}.  

We again model our nodes as a collection of negative point charges 
in $d$ dimensions.  The growth procedure is similar to the previous 
models; we place new nodes randomly in our volume and connect them 
to their $m$ nearest neighbors, where here we define nearest to be 
the Euclidean distance between the nodes.  Now, however, we regard 
the nodes as free to move in a unit radius, $d$-dimensional ball.  
(For $d=1$, the unit ball is the interval $-1\le x \le 1$; for $d=2$, 
it is the region enclosed by the unit circle.)  We assume that the 
ball contains a uniform background positive charge density such that 
the total background charge in the sphere is equal and opposite to 
that of the $N$ network nodes.  As in the sphere model, after adding 
a node with uniform probability density within the unit $d$-dimensional 
ball, we relax the charge configuration to a local energy minimum.  
Here, the relaxation is described by 
\begin{equation} 
\frac{\text{d}\bm{x}_i}{\text{d}t} = \bm{F}_i - N \bm{x}_i 
\end{equation} 
where $\bm{x}_i$ is a $d$-dimensional position vector with respect 
to the center of the ball, $\bm{F}_i$ is as in Eq.\ \eqref{eq:force}, 
and the term $N\bm{x}_i$ is due to the positive charge density.  

Note that, in order to apply Gauss's law for the background charge, 
we have assumed a force law proportional to $r^{-(d-1)}$.  Gauss's 
law, in turn, implies that when $N$ is large, the nodes will be 
approximately uniformly distributed in the ball in order to cancel 
the uniform positive background charge.  Although any repulsive 
force law can, in principle, be used for the sphere model, we chose 
to use the same force law in Sec.\ \ref{sec:spheremodel} in order 
to facilitate comparisons of the results between the two models.

\begin{figure}[t] 
\includegraphics[width=\linewidth]{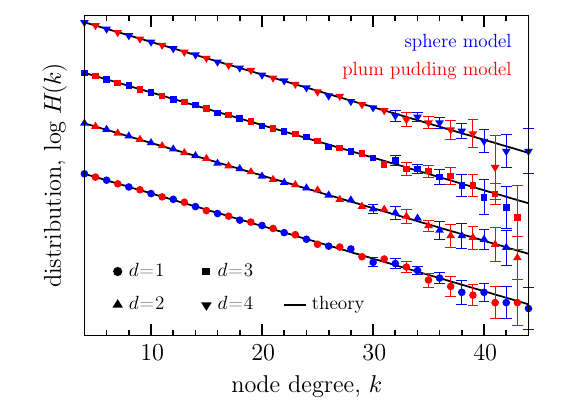} 
\caption{\label{fig:degdist} 
The logarithm of the degree distribution, $\log(H(k))$, versus degree 
$k$, for the sphere network model (blue markers) and the plum pudding 
model (red markers), using $N = 10^4$ and $m=4$.  Data are shown for 
$d=1$ (circles), $d=2$ (triangles), $d=3$ (squares), and $d=4$ (inverted 
triangles).  Since all four cases have nearly identical results, an 
arbitrary linear offset has been used to separate data for visualization.  
Error bars are shown when they exceed the point size.  Solid black lines 
correspond to the theoretical prediction of Eq.\ \eqref{eq:degdist}.} 
\end{figure}

\section{Degree Distribution\label{sec:degdist}}

The distribution of node degrees can be derived analytically, does 
not depend on dimension, and is the same for the sphere and plum 
pudding models.  This can be derived from the fact that, for large 
$N$, the probability that a newly added node will form an edge to 
any particular existing node is $m/N$ for all nodes.  This is because 
existing nodes are distributed approximately uniformly, and new nodes 
are placed randomly according to a uniform probability distribution.  
Here we show that for each considered model, we produce the same 
master equation governing the evolution of the degree distribution 
as that found for the circle model in \cite{ozik2004}.  This master 
equation is not specific to the spatial structure of the network and 
appears, in various forms, in other network models, such as the 
Deterministic Uniform Random Tree of Ref.~\cite{zhang2008topologies}.  

We define $\hat{G}(k,N)$ to be the number of nodes with degree $k$ 
at growth step $N$ (i.e., when the system has $N$ nodes).  When a 
node is added to the network it is initially connected to its $m$ 
nearest neighbors, so upon creation, $k = m$ for each node, meaning 
that $\hat{G}(k,N) = 0$ for $k < m$.  Since each existing node is 
equally likely to be chosen to be connected to the new node, there 
is an $m/N$ probability that any given node will have its degree 
incremented by 1.  Averaging $\hat{G}(k,N)$ over all possible random 
node placements, we obtain a master equation for the evolution of 
$G(k,N)$, the average of $\hat{G}(k,N)$ over all possible randomly 
grown networks, 
\begin{equation} 
\begin{split} 
G(k,N+1) = G(k,N) &- \frac{m}{N}G(k,N) \\ 
                  &+ \frac{m}{N}G(k-1,N) + \delta_{km}, 
\end{split} 
\end{equation} 
\noindent where $\delta_{km}$ is the Kronecker delta function.  The 
first term on the right is the expected number of nodes with degree 
$k$ at growth step $N$.  The second term is the expected number of 
nodes with degree $k$ at growth step $N$ that are promoted to degree 
$k+1$.  The third term is the expected number of nodes with degree 
$k-1$ at growth step $N$ that are promoted to degree $k$.  The last 
term on the right is the new node with degree $m$.  

\begin{figure}[t] 
\includegraphics[width=\linewidth]{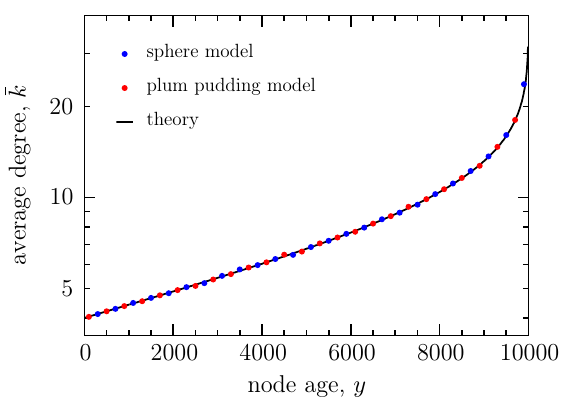} 
\caption{\label{fig:degage} 
The mean degree $\bar{k}(y,N)$ of a node of age $y$ in a network with 
$N$ nodes on semilogarithmic axes.  Data are shown for the sphere network 
model (blue) and plum pudding model (red), both with $d=2$, $m=4$, and 
$N=10^4$.  Errors are smaller than the point size.  The solid black line 
is the theoretical prediction given by Eq.~\eqref{eq:age}.} 
\end{figure} 

It was shown by Ozik et al.\ \cite{ozik2004} that this master equation 
leads to an exponentially decaying degree distribution with an 
asymptotically $N$ invariant form $H(k) = \lim_{N \to \infty} G(k,N)/N$ 
given by Eq.~\eqref{eq:degdist} for $k \geq m$ and $H(k) = 0$ for $k < m$.  
Interestingly, this degree distribution comes only from the growth 
process and the uniform probability of attaching new links to existing 
nodes.  As seen in Fig.~\ref{fig:degdist}, for $m = 4$, $N = 10^4$, with 
$d = 1$, $2$, $3$, or $4$, Eq.~\eqref{eq:degdist} is well satisfied by 
numerical simulations of both models.  

Intuitively, we expect that older nodes in each model will accumulate 
more edges and become network hubs.  The relationship between degree 
and age is also straightforward to investigate in this model.  Here, 
we calculate an expression for the expected degree $\bar{k}(y,N)$ of 
a node that has existed for $y$ growth steps, given that the network 
size is $N$ ($y < N$).  Each node connects to its $m$ nearest neighbors 
upon creation, and the probability of incrementing the degree of the 
node is $m/N$ when the size of the network is $N$.  Thus we obtain 
\begin{equation}\label{eq:age} 
\begin{split} 
\bar{k}(y,N)& = m + m\sum_{n=N-y+1}^{N} \frac{1}{n}\\ 
            & \approx m + m \log \frac{N}{N-y} + \mathcal O\left(\frac{1}{N^2}\right).  
\end{split} 
\end{equation} 
 
Once again, since this derivation uses only the assumption that each 
node has an equal chance each growth step to have its degree incremented, 
the result holds for both of the models discussed here.  This represents 
a specific example of the fact that in dynamically growing networks, 
older nodes are preferentially connected to subsequent nodes, as discussed 
in Ref.~\cite{reallyrandom}.  Numerical simulations in Fig.~\ref{fig:degage} 
demonstrate that Eq.~\eqref{eq:age} is satisfied for both models.  For 
simplicity, results are only presented for $d=2$, but Eq.~\eqref{eq:age} 
has no dependence on the embedding space, and thus holds for other 
dimensions as well.

\begin{figure}[t] 
\includegraphics[width=\linewidth]{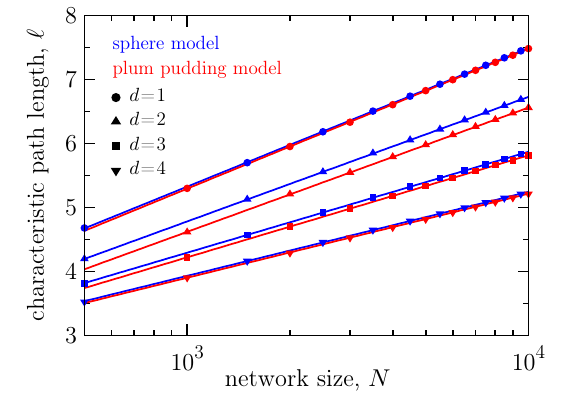} 
\caption{\label{fig:pathlength} 
The characteristic graph path length ($\ell$) versus network size $N$ 
on semilogarithmic axes for the sphere model (blue markers) and the 
plum pudding model (red markers).  The path length shows the desired 
scaling, $\ell \sim \log N$.  Results are shown for $d=1$ (circles), 
$2$ (triangles), $3$ (squares), and $4$ (inverted triangles), all using 
$m=4$.  Errors are smaller than the point size.  In general, the average 
shortest path is shorter for higher dimensions $d$ in both models.} 
\end{figure}

\section{Path Length and Clustering Coefficient\label{sec:pathclust}} 

For the sphere and plum pudding network models, we find numerically 
that the the average shortest path length $\ell$ scales logarithmically 
with the network size $N$, that is, $\ell \sim \log N$.  See Fig.\ %
\ref{fig:pathlength}.  The scaling $\ell \sim \log N$ is expected 
because as the network grows in size, the older nodes are pushed 
apart by the repulsive force, thus leaving bridges across the network 
that span a significant geographic distance.  These long range links 
serve to connect spatially separated regions of highly interconnected 
nodes, dramatically reducing the shortest path length between any two 
nodes in the network.  At each growth step only geographically local 
connections are made, but due to the dynamic nature of the nodes' 
spatial positions, each growth step can make existing links longer 
in physical space, thus building bridges across the network.  

We see from Fig.~\ref{fig:pathlength} that, for given values of $N$ and 
$m$, the characteristic path length decreases with $d$ and is shorter 
than that of the corresponding one-dimensional case (the original circle 
network model).  One possible explanation for this is that in a higher 
dimensional space, it is easier to separate existing nodes by placing 
a new node, because nodes can move around one another, making it easier 
for short-range links to be stretched into shortcuts as the network 
grows.  This is in contrast to the circle network model, in which each 
node is forever locked between its two original spatial neighbors until 
a new node is placed directly between them.  

\begin{figure}[t] 
\includegraphics[width=\linewidth]{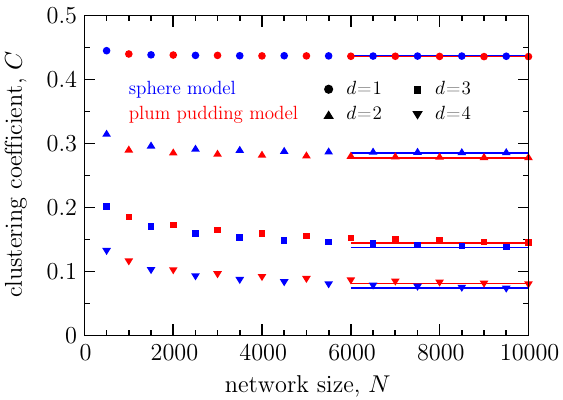} 
\caption{\label{fig:clustN} 
Clustering coefficient $C$ versus network size $N$ for the sphere 
network model (blue markers) and plum pudding model (red markers), 
both with $m=4$.  Results are shown for $d=1$ (circles), $2$ 
(triangles), $3$ (squares), and $4$ (inverted triangles).  Errors 
are smaller than the point size.  Horizontal lines are drawn 
through the last data point in each series.} 
\end{figure} 

\begin{figure}[t] 
\includegraphics[width=\linewidth]{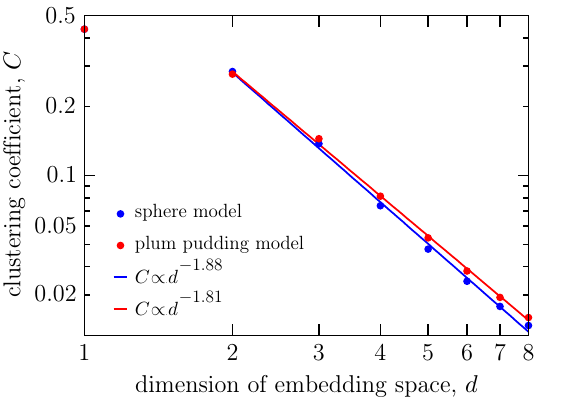} 
\caption{\label{fig:clustdim} 
Asymptotic clustering coefficient $C$ versus dimension of embedding 
space $d$ for the sphere network model (blue) and plum pudding model 
(red) on logarithmic axes with $m=4$.  Solid black lines are given 
by $C \propto d^{-\beta}$, $\beta$ was obtained from a least-squares 
fit to the data for $d \geq 2$.  The resulting value of $\beta$ was 
$1.88$ for the sphere model and $1.81$ for the plum pudding model.} 
\end{figure}

For both models, we also find that the clustering coefficient $C$ 
is nonzero for large $N$, but depends on the dimension of the 
embedding space.  Results for the clustering coefficient $C$ versus 
the number of nodes $N$ with $d=1$, $2$, $3$, and $4$, using $m=4$, 
are displayed in Fig.~\ref{fig:clustN}.  Horizontal lines are drawn through 
the last point in each series.  For the values of $d$ shown, we see 
that as $N$ increases, there is an initial decrease of $C$ for $N<1,000$, 
but the $N$ variation appears to effectively cease with increasing 
$N$.  These results are consistent with an $N \to \infty$ asymptotic 
value close to the value at $N=10,000$.  Assuming this to be the case, 
values for the large-$N$ clustering in the sphere model are given as 
follows: for $d=1$, $C \approx 0.44$; for $d=2$, $C \approx 0.28$; 
for $d=3$, $C \approx 0.14$; and for $d=4$, $C \approx 0.07$.  Values 
for the plum pudding model are similar.  Higher dimensional cases that 
we have examined ($d=5$--$9$) follow the same pattern.  More 
specifically, we find that for a given value of $m$, the clustering 
coefficient decays algebraically with dimension, $C \sim d^{-\beta_m}$.  
In Fig.~\ref{fig:clustdim}, we show that $\beta_4 = 1.88$ for the 
sphere model, while for larger values of $m$ we find $\beta_m$ 
decreases, but remains positive.  

Thus we find that that both the sphere and plum pudding network models 
lead to networks exhibiting the small-world property, and their 
behaviors are similar.  Thus spatial topology does not appear to be 
of great importance to the properties of the resulting network.  Node 
addition, local edge formation, and relaxation towards uniform density 
appear to be sufficient for the occurrence of the small-world property.  
On the other hand, although the degree distribution does not depend on 
dimension (Sec.\ \ref{sec:degdist}), an important result is that other 
network properties such as the clustering coefficient show a relatively 
strong dependence on dimensionality (Figs.\ \ref{fig:pathlength}--%
\ref{fig:clustdim}).  

It is worth noting that the $d=1$ case of the plum pudding model does 
not reduce to the circle network model, but the two nonetheless show 
similar behavior, because they differ only at the boundary $x=\pm 1$.  
In both models, the relative spatial ordering of nodes is preserved, 
and the equilibrium case has perfectly uniform inter-nodal spacing, 
as opposed to models with $d \geq 2$, in which the concept of linear 
ordering is absent.  In addition, when $d \geq 2$ in both the sphere 
and plum pudding models, exact, global regular-lattice positioning 
is not possible.  For example, for $N \gg 1$ charges on a sphere with 
$d=2$, it is known that the equilibrium positioning on much of the 
area of the sphere is locally similar to a triangular lattice, but 
the sphere's curvature leads to point and line defects in the lattice 
\cite{nelson}.  Thus a natural question is whether the $d=1$ cases 
might have special properties in common that deviate from those for 
$d \geq 2$.  It can be seen in Fig.~\ref{fig:clustdim} that one such 
property is that both $d=1$ cases do not follow the same scaling 
trend in clustering that we find for higher dimensions.

\section{Discussion} 

We have explored two models which generate networks with small-world 
features through local geographic attachment and growth, without 
direct formation of long-distance links.  By allowing nodes to move 
in space, the initially formed local links can become long-range, 
thus providing a mechanism for how small-world networks can emerge 
from a growing collection of dynamically interacting and locally 
constrained vertices.  Both models show similar behavior for the 
degree distribution, characteristic path length, and clustering 
coefficient.  The qualitative similarity between the networks 
generated by the two models indicates that the small-world features 
are determined by geographic attachment and growth, not the 
topological features of the embedding space.  However, the 
quantitative values of measures characterizing these features 
can depend on dimension; higher-dimensional spaces yield shorter 
path lengths but less clustering.  

These findings may offer insight into the origin of small-world 
features in diverse growing networks, such as power grids and 
networks of neurons.  In such systems, network growth may be an 
appropriate mechanism for the emergence of small-world features.  
We also speculate that similar ideas may explain the small-world 
property in some non-spatial networks, such as the world wide web, 
by replacing the physical space used in our model with a more 
abstract space of content (i.e., the location of a node represents 
the topic or purpose of a website, and websites link to other 
websites which have similar content).  

We hope these findings will generate renewed interest in spatial 
networks with dynamically located nodes and in the role that 
growth plays in the development of important network features.  

\textit{Acknowledgements:} This work was supported by the National 
Science Foundation under grant number PHY-1156454 and by the Army 
Research Office under grant W911NF-12-1-0101.  

\bibliography{grownet.bib} 

\end{document}